 \documentclass[aps,preprint,preprintnumbers, amsmath, amssymb, prb]{revtex4-1}
\usepackage{times}
\usepackage{graphicx}
\usepackage{psfrag}
\usepackage{ae}
\usepackage{amsmath,amssymb}
\usepackage[usenames]{color}
\usepackage{float}

\begin{document}
\date{\today}

\author{Jos\'e Rafael Bordin} 
\email{josebordin@unipampa.edu.br}
\affiliation{Campus Ca\c capava do Sul, Universidade Federal
do Pampa, Caixa Postal 15051, CEP 96570-000, 
Ca\c capava do Sul, RS, Brazil}

\author{Leandro B. Krott} 
\email{leandro.krott@ufrgs.br}
 \affiliation{Programa de P\'os-Gradua\c c\~ao em F\'{\i}sica, Instituto 
 de F\'{\i}sica, Universidade Federal
 do Rio Grande do Sul, Caixa Postal 15051, CEP 91501-970, 
  Porto Alegre, RS, Brazil}

\author{Marcia C. Barbosa} 
\email{marciabarbosa@ufrgs.br}
\affiliation{Instituto 
de F\'{\i}sica, Universidade Federal
do Rio Grande do Sul, Caixa Postal 15051, CEP 91501-970, 
Porto Alegre, RS, Brazil}

\title{Surface Phase Transition in Anomalous Fluid in Nanoconfinement}


\begin{abstract}
We explore by molecular dynamic simulations
the thermodynamical behavior of an anomalous fluid confined 
inside rigid and flexible nanopores. 
The fluid is modeled by a  two length scale potential.
In the bulk this system exhibits the density and 
diffusion anomalous behavior observed in liquid water.
We show that the anomalous fluid confined inside 
rigid and flexible nanopores forms layers. As
the volume of the nanopore is decreased the rigid surface 
exhibits three consecutive
first order phase transitions associated
with the change in the number of layers. These
phase transitions are not present for flexible confinement.
Our results indicate that the nature of confinement
is relevant for the properties of the confined liquid what
suggests that confinement in carbon nanotubes should
be quite different from confinement in biological channels.

\end{abstract}


\maketitle
\section{Introduction}

Most liquids contract on cooling and 
diffuse faster as the density is 
decreased. This is not the case 
of the anomalous liquids in which the density exhibits
a maximum at constant pressure and the diffusion coefficient increases under 
compression~\cite{URL}. 
These anomalous fluids include water~\cite{Ke75,An76,Pr87},
Te~\cite{Th76}, Ga,
Bi~\cite{Handbook}, Si~\cite{Sa67,Ke83}, $Ge_{15}Te_{85}$~\cite{Ts91},  liquid 
metals~\cite{Cu81} and graphite~\cite{To97}.
Computer simulations for silica~\cite{An00,Sh02,Sh06}, silicon~\cite{Sa03}  
and $BeF_2$~\cite{An00} also show the presence of 
thermodynamic anomalies~\cite{Pr87}. In addition to the presence
of a maximum of density in constant pressure, 
silica~\cite{Sa03,Sh02,Sh06,Ch06}, silicon~\cite{Mo05} 
and water~\cite{Ne01, Ne02a}  exhibit  a maximum in the 
diffusion coefficient at constant temperature.

Classical all-atom models such as  SPC/E~\cite{spce}, TIP4P-2005~\cite{Ab05} 
and TIP5P~\cite{Ma00} for
water, sW~\cite{St85} for silicon or BKS~\cite{Be90} for silica
have been employed  to reproduce
quantitatively these anomalous properties of these
materials. However,  coarse-grained potentials are an interesting 
tool able to identify what is the common structural property in these
fluids that make them anomalous. The effective potentials 
derived in these coarse grained models
are analytically more tractable and
also computationally less expensive, what allow for studying a
very large systems  and complex mixtures.  Several effective models 
have been 
proposed~\cite{Ja98,Xu06,Ya05, Xu05,Oliveira06a,Ca05,Wi02,Al09,Fo08,Fr01,Fr07}.
They reproduce the thermodynamic, structural and dynamic anomalies present in water and in
other anomalous liquids.
The common ingredient in these 
potentials is that the particle-particle interaction is modeled 
through core-softened potentials formed by two length scales, one repulsive shoulder and
an attractive well~\cite{Oliveira06a,Oliveira06b,Barraz09,Silva10}. These
competition leads to the density and diffusion anomalies. 

In addition to the bulk properties,
nanoconfinement of anomalous liquids  has been attracting attention not only
due to its applications but also due to the new physics
observed in these systems~\cite{Ma05,Holt06,Wh08}. Fluids confined in
carbon nanotube exhibit formation of layers, crystallization
of the contact layer~\cite{Cu01, JCP06} and a superflow not
present in macroscopic confinement~\cite{Jakobtorweihen05,Chen06b,Qin11}.

In the particular case of  water confined in nanopores, the
pore size has significant influence on the freezing and melting temperatures of
water~\cite{Er11,De10,Ja08,Mo97}. The crystallization in these systems 
is not uniform and
the confined ice shows different characteristics when
compared with  the bulk 
ice~\cite{Kastelowitz10}.
Hydrophobic~\cite{Mao02,Ackerman03,Qin11,Khademi11}
and hydrophilic~\cite{Lee12} confinements also induce different effects in 
the layering, density and flow of water.

Atomistic studies of nanoconfinement of 
water  show  another
property:
confined systems exhibit a phase transition not observed in the bulk system.
SPC/E model confined between atomically smooth plates~\cite{Giovambattista09, Lo09}
and TIP4P water inside nanotubes~\cite{Koga01} shows a first order phase 
transition between a bilayer liquid (or ice) and a 
trilayer heterogeneous fluid.
These studies, however,  has been restricted to
rigid nanotubes. The flexibility of the nanochannel~\cite{Jakobtorweihen05,Chen06b}
and of biological ionic channels~\cite{Noskov04, Beckstein04,Allen04, Chiu91} show
properties different from the behavior observed in confinement by rigid walls. These 
studies, however, do not highlight the physical reason behind the differences
between rigid and flexible confinement.

Acknowledging that coarse graining potentials would be a suitable
tool to test how the flexibility would affect the properties of confined
anomalous liquids. Recently it was shown that the density and
diffusion anomalies disappears as the channel or nanopore become
flexible~\cite{Krott13b}.

In this paper we explore the differences in the  layering and in the
surface phase transitions for anomalous
fluids confined by  both 
rigid and flexible 
nanotubes. We show that
the surface crystallization observed
in rigid carbon nanotubes should not be expected
in flexibly biological channels.  The fluid is modeled using a two length
scale potential. This coarse grained potential
exhibits the thermodynamic, dynamic and structural anomalous behavior
observed in anomalous fluids in bulk~\cite{Oliveira06a,Oliveira06b} 
and in confinement~\cite{Krott13a,Krott13b,Krott14a,Bordin12b,Bordin13a}. The 
formation of layers and its 
relation with the first order phase transition are analyzed. The paper is 
organized as follows: in Sec. II we introduce the model
and describe the methods and simulation details; the results are given in Sec. III; and 
in Sec. IV we present our conclusions.

\section{The Model and the Simulation Methodology}

\begin{figure}[ht]
\begin{center}
\includegraphics[width=8cm]{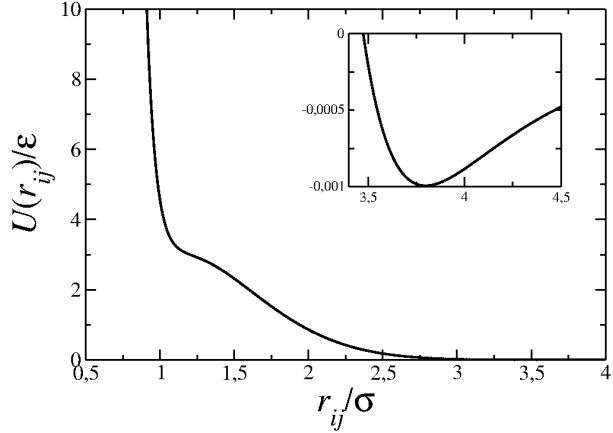}
\end{center}
\caption{Interaction potential between anomalous fluid particles pair as 
function of their separation. Inset: zoom over the small
attractive part of the interaction.}
\label{fig1}
\end{figure}

The fluid is modeled as spherical-symmetric particles, with diameter $\sigma$ and mass $m$.
The particles interact through the three 
dimensional core-softened potential~\cite{Oliveira06a}
\begin{equation}
\frac{U(r_{ij})}{\varepsilon} = 4\left[ \left(\frac{\sigma}{r_{ij}}\right)^{12} 
-\left(\frac{\sigma}{r_{ij}}\right)^6 \right]
 + u_0 {\rm{exp}}\left[-\frac{1}{c_0^2}\left(\frac{r_{ij}-r_0}{\sigma}\right)^2\right]
 \label{AlanEq}
 \end{equation}
where $r_{ij} = |\vec r_i - \vec r_j|$ is the distance between two fluid 
particles $i$ and $j$.
This potential has two contributions. The first parcel is the standard $12-6$
Lennard-Jones (LJ)
potential~\cite{AllenTild} and the second term  is a Gaussian
centered at $r_0/\sigma = 0.7$, with depth $u_0\varepsilon = 5.0$ and width 
$c_0\sigma = 1.0$.
With these parameters, the equation~\ref{AlanEq} represents a two length scale 
potential, with one scale 
at  $r_{ij}\approx 1.2 \sigma$, where the 
force has a local minimum, and the other scale at  $r_{ij} \approx 2 \sigma$, where
the fraction of imaginary modes has a local minimum~\cite{Oliveira10}, as 
shown in figure~\ref{fig1}.
The fluid-fluid interaction, equation~(\ref{AlanEq}), has a cutoff radius 
$r_{\rm cut}/\sigma = 3.5$.
Despite the mathematical simplicity of the model, this fluid exhibits 
the thermodynamic, dynamic and structural anomalies present 
in bulk water~\cite{Oliveira06a, Oliveira06b}
and a water-like behavior when confined between 
plates~\cite{Krott13a, Krott13b, Krott14a} or inside 
hydrophobic nanotubes~\cite{Bordin12b, Bordin13a}.

\begin{figure}[ht]
\begin{center}
\includegraphics[width=8cm]{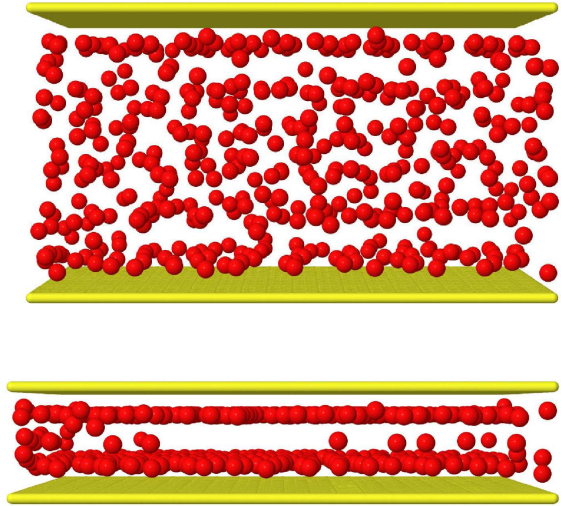}
\end{center}
\caption{Snapshot of the simulation box for system for large (up) and 
narrow (down) nanopores.}
\label{fig2}
\end{figure}

The nanopore was modeled using two flat parallel walls, with 
fixed dimension $L\times L$,
where $L = 40\sigma$, separated by a distance $L_z$. In the 
figure~\ref{fig2} we show the snapshot of the system for two
distinct configurations, one with a large $L_z$, where 
the fluid shows a bulk-like behavior, and the other a highly
confined fluid. The fluid-wall interaction is purely repulsive, and 
was represented by
the Weeks-Chandler-Andersen (WCA)~\cite{WCA71} potential,

\begin{equation}
\label{LJCS}
U^{\rm{WCA}}(z_{ij}) = \left\{ \begin{array}{ll}
U_{{\rm {LJ}}}(z_{ij}) - U_{{\rm{LJ}}}(z_c)\;, \qquad z_{ij} \le z_c\;, \\
0\;, \qquad \qquad \qquad \qquad \quad z_{ij}  > z_c\;.
\end{array} \right.
\end{equation}
Here, $U_{{\rm {LJ}}}$ is the standard $12-6$ LJ potential, included in the 
first term of equation~(\ref{AlanEq}),
and $z_c = 2^{1/6}\sigma$ is the cutoff for the WCA potential. Also, the 
term $z_{ij}$ measures the distance
between the wall at $j$ position and the $z$-coordinate of the fluid particle $i$.

Two distinct scenarios were studied: rigid and 
flexible walls. In the first case, the nanopore  walls positions
were fixed and standard $NVT$ Molecular
Dynamic simulations were performed. 
The temperature control was
obtained with the Nos\`e-Hoover thermostat, with a coupling parameter 
$Q = 2.0$. The pressure in the 
$z$ direction, $p_z$, was computed by the virial expression in the 
direction of the confinement ($z$)~\cite{Zangi00}, namely
\begin{equation}
p_z = \rho k_b T + \frac{1}{V}\langle W_t \rangle\;,
\end{equation}
where 
$$\langle W_t \rangle = -\sum_1^N \sum_{j>1}^N \frac{z_{ij}^2}{r_{ij}} \frac{\partial U}{\partial r}\;,
$$
and $U(r_{ij})$ is the interaction potential between two particles separated by a distance $r_{ij}$,
and $z_{ij}$ is the $z$-component of the distance.

In a second scenario flexible walls were studied. In this case, 
MD simulations were performed
at constant number of particles and perpendicular pressure and 
temperature ($Np_{z}T$ ensemble). The pressure was fixed
using the Lupkowski and van Smol method~\cite{LupSmol90}. 
The walls had  translational freedom in the $z$-direction,
acting like a piston in the fluid, and a constant force controls the 
pressure applied in the confined direction.
In this scenario, the resulting force in a fluid particle is given by
\begin{equation}
 \vec F_R = -\vec\nabla U + \vec F_{iwA}(\vec r_{iA}) + \vec F_{iwB}(\vec r_{iB})\;,
\end{equation}
where  $\vec F_{iwA(B)}$ indicates the interaction 
between the particle $i$ and the piston $A(B)$. 
Once the walls are non-rigid and time-dependent, we have to solve the equations of motion
for $A$ and $B$,
\begin{equation}
 m_w\vec a_A = p_{z}S_w\vec n_A - \sum_{i=1}^N \vec F_{iwA}(\vec r_{iA})
\end{equation}
and
\begin{equation}
 m_w\vec a_B = p_{z}S_w\vec n_B - \sum_{i=1}^N \vec F_{iwB}(\vec r_{iB})\;,
\end{equation}
respectively, where $m_w$ is the piston mass, $p_{z}$ is the applied pressure in the 
system, $S_w$ is the piston area and $\vec n_A$ is an unitary vector 
in positive $z$-direction, while $\vec n_B$ is a negative unitary vector. Both 
pistons ($A$ and $B$) have mass $m_w=m$, width $\sigma$ and area equal to $S_w = L^2$.

For the rigid nanopore system, the temperature was varied from 
$k_BT = 0.05$ to $k_BT = 1.00$, and the
plate separation from $L_z = 4.00\sigma$ to $L_z = 10.00\sigma$, while 
for systems with flexible nanopores the
temperature was varied from $k_BT = 0.10$ to $k_BT = 1.00$, and the 
perpendicular pressure from
$p_{z}\sigma^3/\epsilon = 0.075$ to $p_{z}\sigma^3/\epsilon = 6.00$. In both cases 
the simulations where performed with $N = 1000$
particles. Standard periodic boundary conditions where applied in the non-confined 
directions.
Five independent runs were performed to evaluate the properties of the 
confined fluid. Each individual 
simulation consists of $1\times10^6$ equilibration steps and $3\times10^6$ steps
for production, with a time step $\delta t = 0.0025$, in LJ units. 

In order to define the fluid characteristics in contact with the nanopore walls, 
the structure of the fluid contact layer was analyzed using the radial distribution 
function $g_{||}(r_{xy})$, defined as

\begin{equation}
\label{gr_lateral}
g_{\parallel}(r_{xy}) \equiv \frac{1}{\rho ^2V}
\sum_{i\neq j} \delta (r-r_{ij}) \left [ \theta\left( \left|z_i-z_j\right| 
\right) - \theta\left(\left|z_i-z_j\right|-\delta z\right) \right].
\end{equation}
where the Heaviside function $\theta (x)$ restricts the sum of particle pair in a 
slab of thickness $\delta z = 1.0$ close to the wall.

The physical quantities in this paper are depict in LJ units~\cite{AllenTild},
\begin{equation}
\label{red1}
r^*\equiv \frac{r}{\sigma}\;,\quad \rho^{*}\equiv \rho \sigma^{3}\;, \quad 
\mbox{and}\quad t^* \equiv t\left(\frac{\epsilon}{m\sigma^2}\right)^{1/2}\;,
\end{equation}
for distance, density of particles and time , respectively, and
\begin{equation}
\label{rad2}
p^*\equiv \frac{p \sigma^{3}}{\epsilon} \quad \mbox{and}\quad 
T^{*}\equiv \frac{k_{B}T}{\epsilon}
\end{equation}
for the pressure and temperature, respectively. 
Since all physical quantities are defined in reduced units in this paper, the $^*$ will be omitted in the 
results discussion. 

In the simulations with flexible nanopores, the 
mean variation in the system size induced by the wall fluctuations are smaller than 
$2\%$. Data errors smaller than the data points are not shown.

\section{Results and Discussion}

\begin{figure}[H]
\begin{center}
\includegraphics[width=10cm]{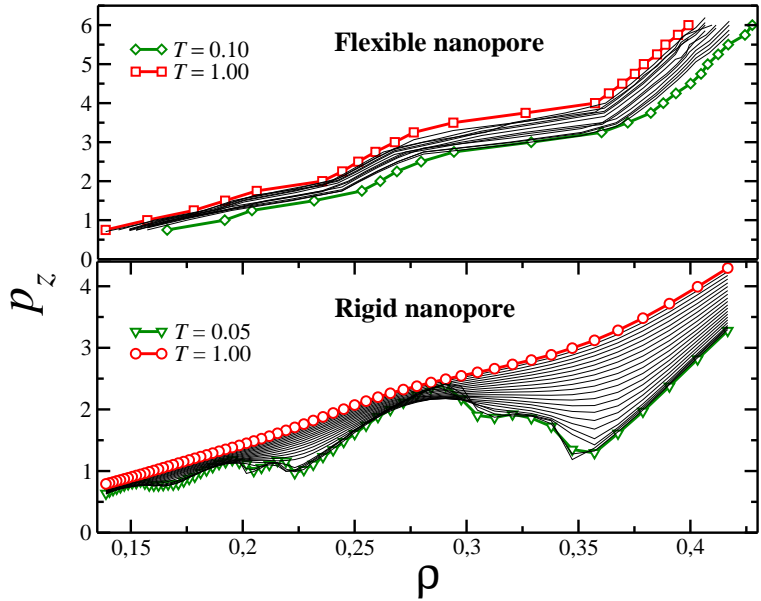}
\end{center}
\caption{
Pressure in the confined direction as function of the fluid density 
for flexible and rigid nanopores
for several values of temperature. For simplicity, only 
the higher and lower value 
of $T$ for the isotherms is detached. A non-monotonic behavior 
was observed when the anomalous fluid is 
confined inside rigid nanopores. All the quantities are in reduced units.}
\label{fig3}
\end{figure}

In order to understand the thermodynamical properties of the anomalous fluid under 
confinement,
the $p_{z}\times\rho$ phase diagram is analyzed for the cases with flexible or 
rigid nanopores.
Figure~\ref{fig3} illustrates the pressure in the confined
direction versus density phase diagram for various temperatures. For the flexible 
nanopores, all the isotherms show a monotonic behavior.
This suggests that while the fluid between the plates changes
its configuration between different layer arrangements the wall
continuously and no phase transition at the wall is observed.
For rigid nanopores, however, the pressure
versus density is a monotonic function for isochores  above  $T_{c3}=0.45$.
Below this temperature, a non-monotonic behavior is observed. 
The isotherms for $T<T_{ci}$ show a van der Waals loop, characteristic of a first
order  phase transition. 

\begin{figure}[H]
\begin{center}
\includegraphics[width=9cm]{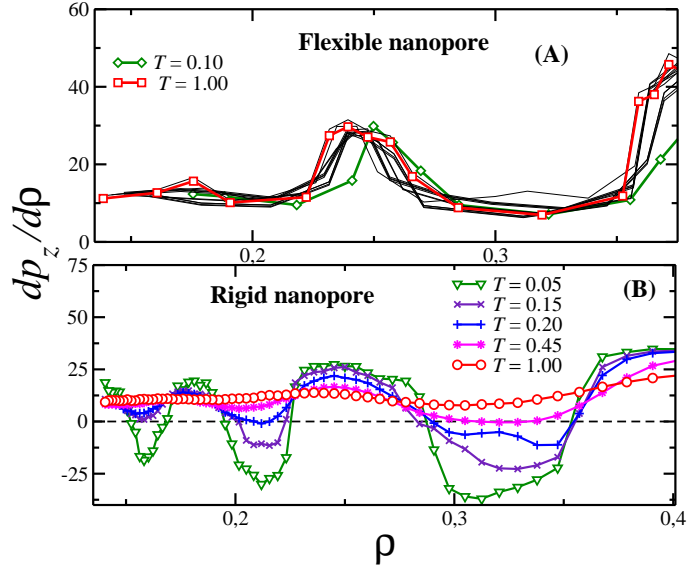}
\end{center}
\caption{Pressure derivative versus density for different isotherms for (A) flexible 
or (B) rigid nanopores. In the rigid case, the inflection in $dp_{z}/d\rho=0$ indicates the 
presence of a first order phase transition.}
\label{fig4}
\end{figure}
\begin{figure}[H]
\begin{center}
\includegraphics[width=9cm]{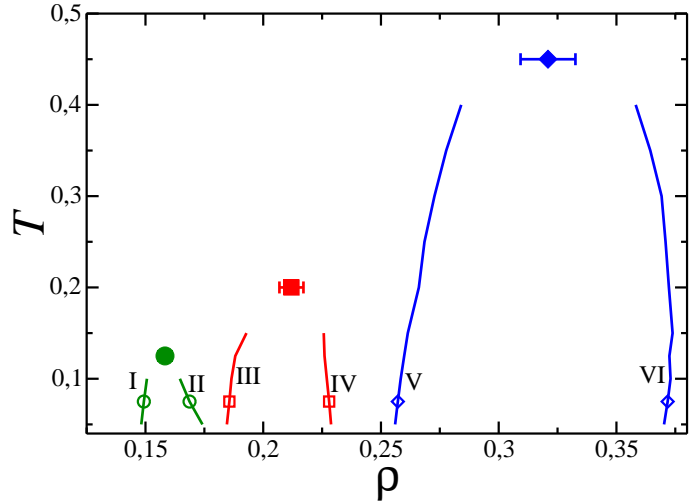}
\end{center}
\caption{ Temperature versus 
density phase diagram for 
the three coexistence regions and critical points: $(T_{c1}=0.125$, $p_{z,c1}=0.782$, $\rho_{c1}=0.1583)$ (sphere), 
$(T_{c2}=0.2$, $p_{z,c2}=1.1704$, $\rho_{c2}=0.212)$ (square) 
and $(T_{c3}=0.45$, $p_{z,c3}=2.235$, $\rho_{c3}=0.321)$ (diamond).The points
I, II, III, IV, V and VI illustrate the coexistence densities at $T=0.075$ 
namely $\rho_I$, $\rho_{II}=$m
$\rho_{III}$, $\rho_{IV}=$, $\rho_V=$ and $\rho_{VI}$.}
\label{fig5}
\end{figure}

For fluids confined 
inside flexible nanopores the pressure derivative with respect
to density  is always positive as 
illustrated by
figure~\ref{fig4}(A). For rigid confinement, the 
derivative is positive only for isochores  $T > T_{c3}$.
Below this threshold the function becomes negative
for various densities as illustrated in figure~\ref{fig4}(B).
This figure identify three first order phase transitions.
The densities of the coexistence phases can
be obtained by Maxell construction. These three coexistence
regions end in  three
critical points that can
be located by computing the second derivative $d^2p/d\rho^2=0$.
The coexisting phases and the three critical points are illustrated by symbols in the
isochores in  figure~\ref{fig5}.

Before discussing the characteristics 
of the fluid and of the phase transition at the wall,
we address the question of why  the  thermodynamical behavior of
flexible nanopores should be different
from the case of rigid nanopores. In the rigid case, the walls only contribute to the 
enthalpic part of the free energy while in the flexible case, the walls vibrations 
constantly shake the fluid particles near the wall, increasing also the entropic part 
of the free energy. While the minimization of the wall-particle and particle-particle 
energy leads to an ordered structure, the entropic contribution from the wall 
disrupts this organization. Therefore, only in the case of rigid walls an 
ordered structure at the wall should be expected. 
Consequently, since the central layers are not affected by the 
wall movement, we can understand the differences between the 
thermodynamical behavior of the confined fluid within rigid
and flexible walls by
analyzing the properties of the layers 
in contact with the walls.
We will refer to this layer as contact layer.

\begin{figure}[H]
\begin{center}
\includegraphics[width=8cm]{Fig6A.eps}
\includegraphics[width=8cm]{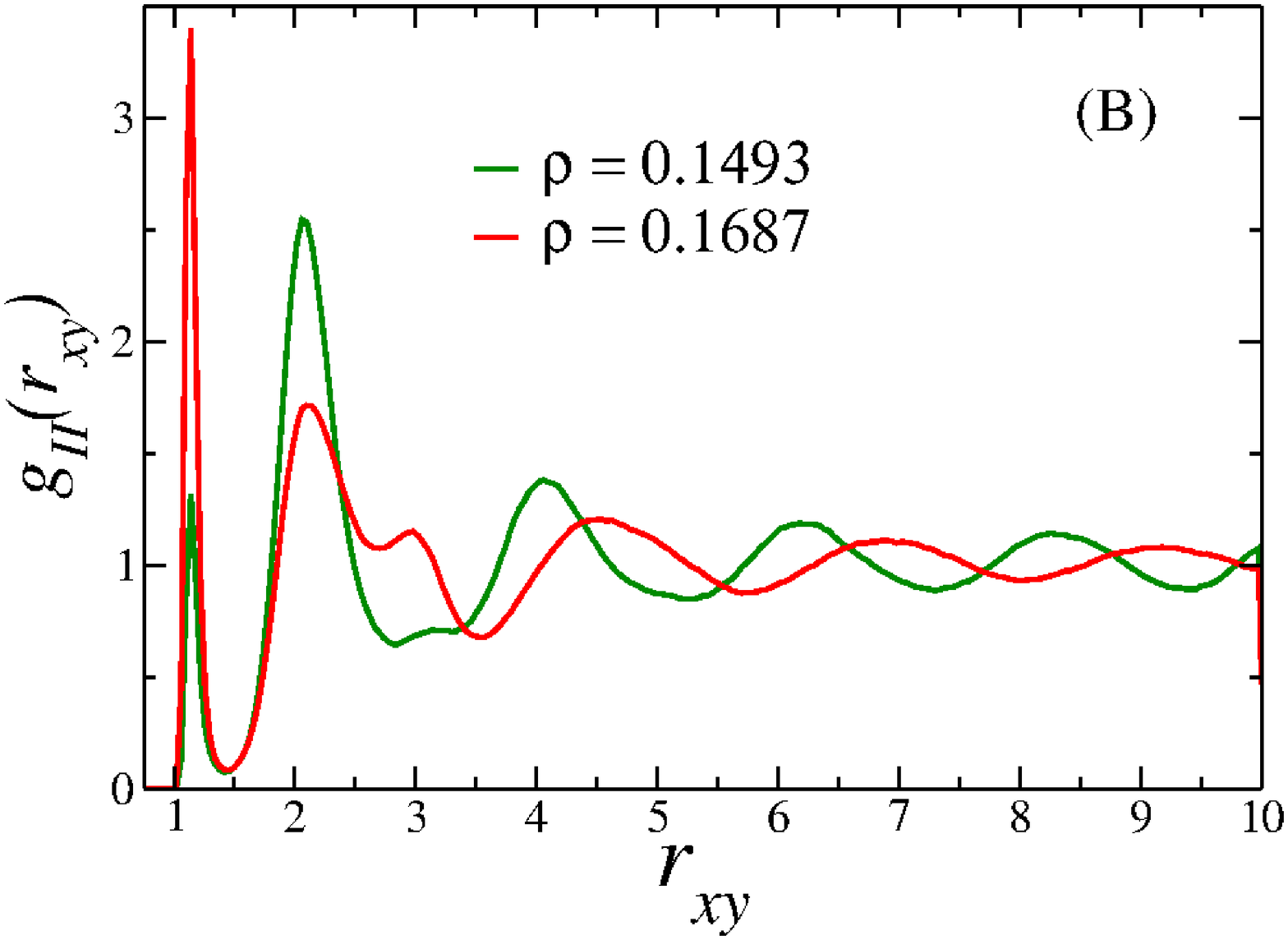}
\includegraphics[width=8cm]{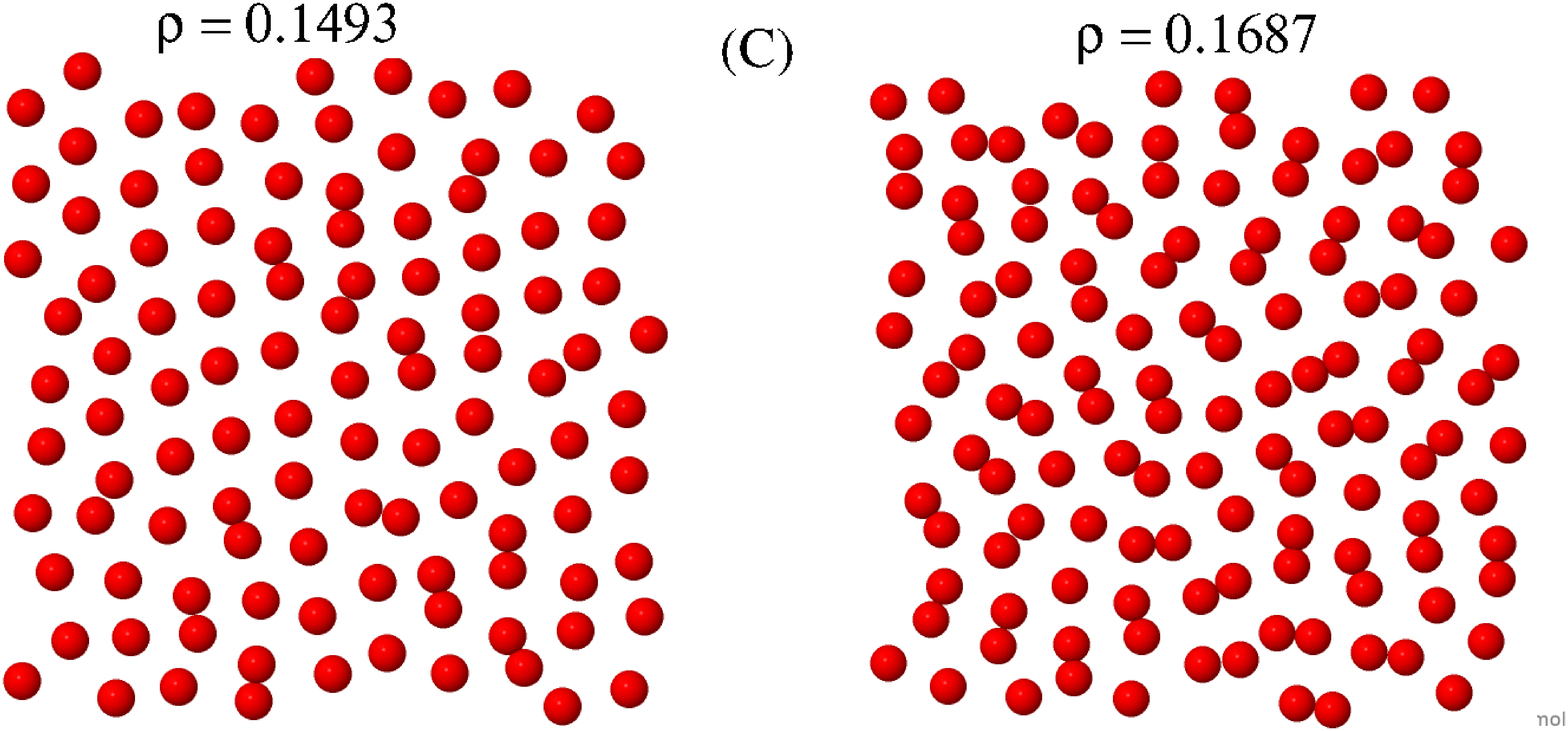}
\end{center}
\caption{Density histogram (A), radial distribution function (B) and snapshots (C) for 
the anomalous fluid confined inside a rigid nanopore at $T = 0.075$
and in the points I and II from figure~\ref{fig5}.}
\label{fig6}
\end{figure}

\begin{figure}[H]
\begin{center}
\includegraphics[width=8cm]{Fig7A.eps}
\includegraphics[width=8cm]{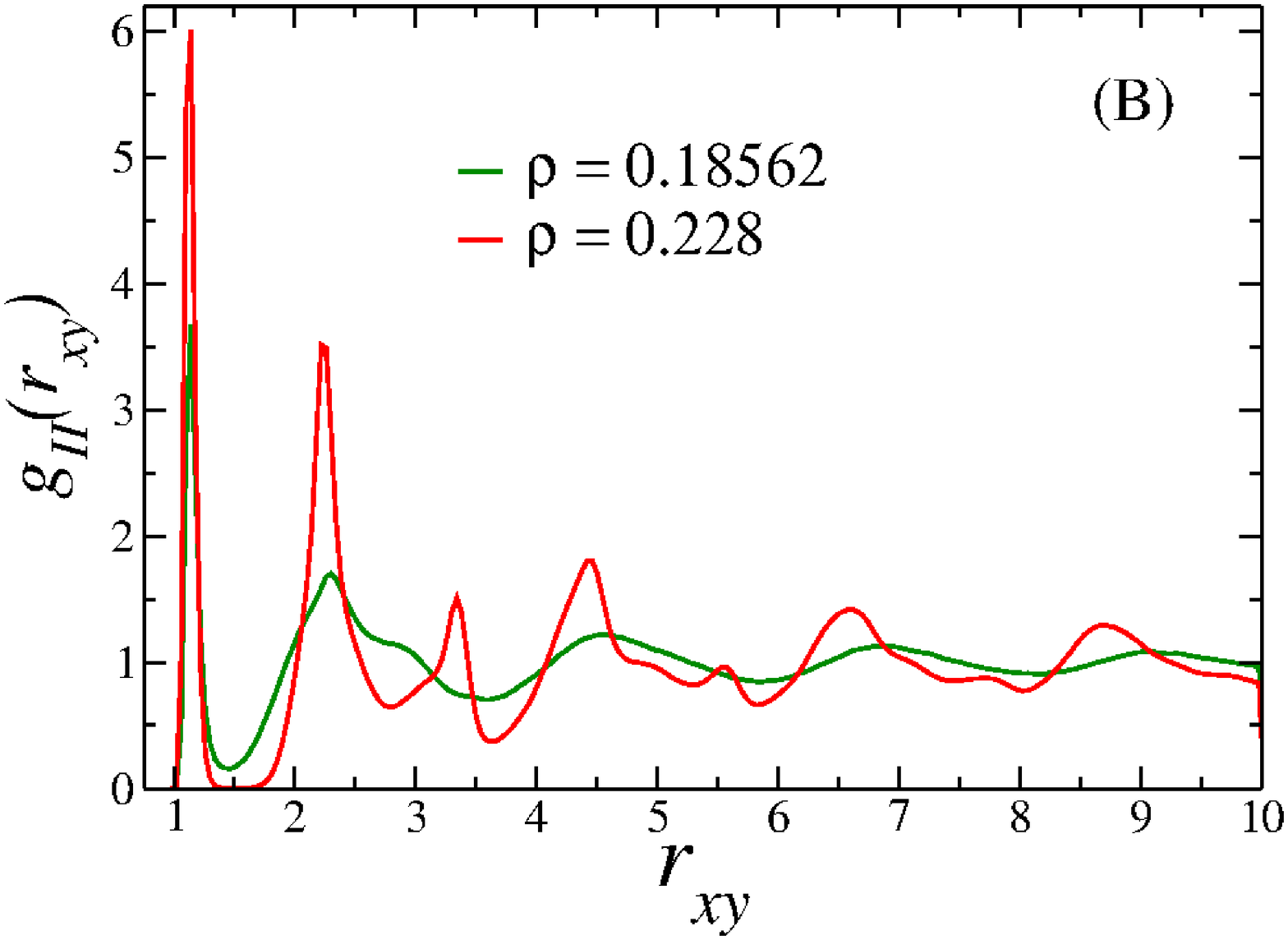}
\includegraphics[width=8cm]{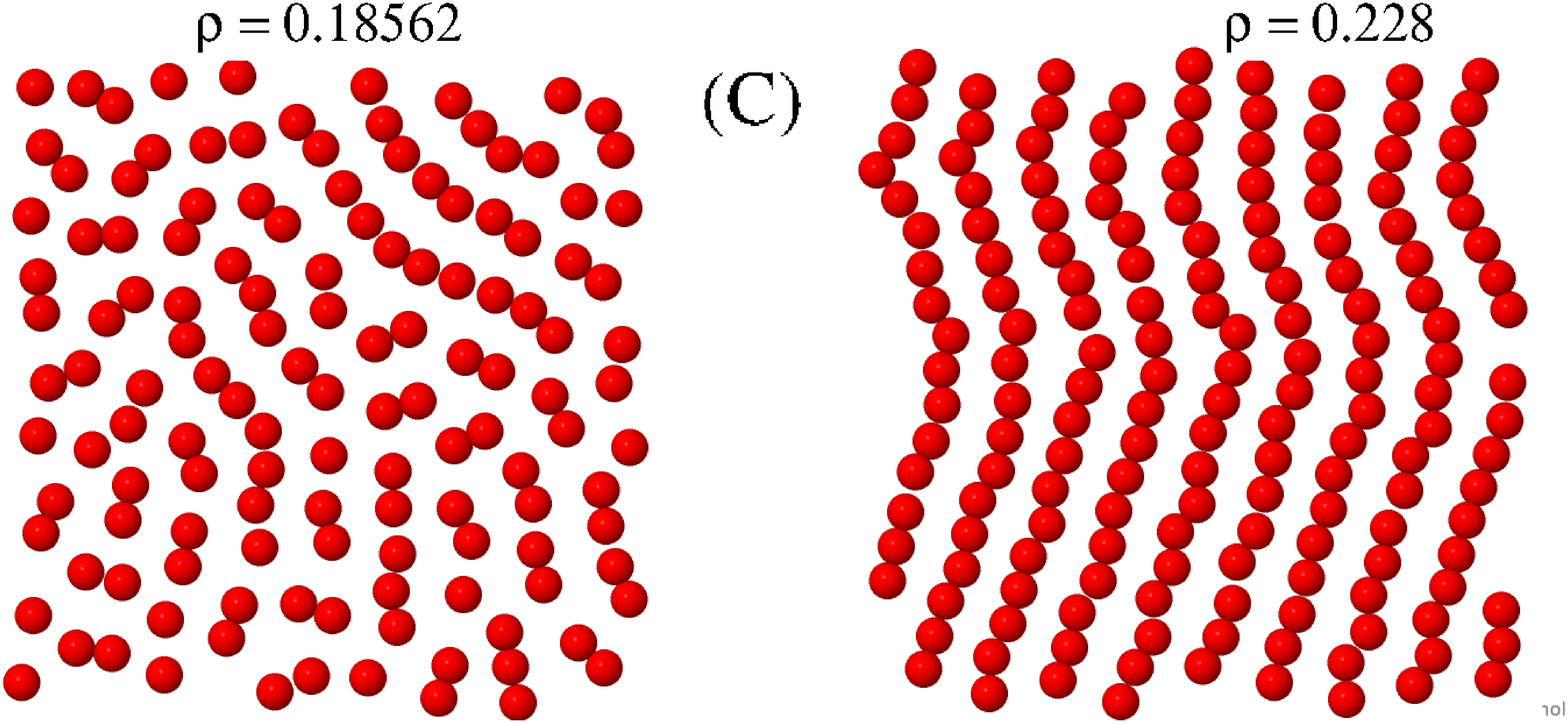}
\end{center}
\caption{Density histogram (A), radial distribution function (B) and snapshots (C) for 
the anomalous fluid confined inside a rigid nanopore at $T = 0.075$
and in the points III and IV from figure~\ref{fig5}.}
\label{fig7}
\end{figure}

\begin{figure}[H]
\begin{center}
\includegraphics[width=8cm]{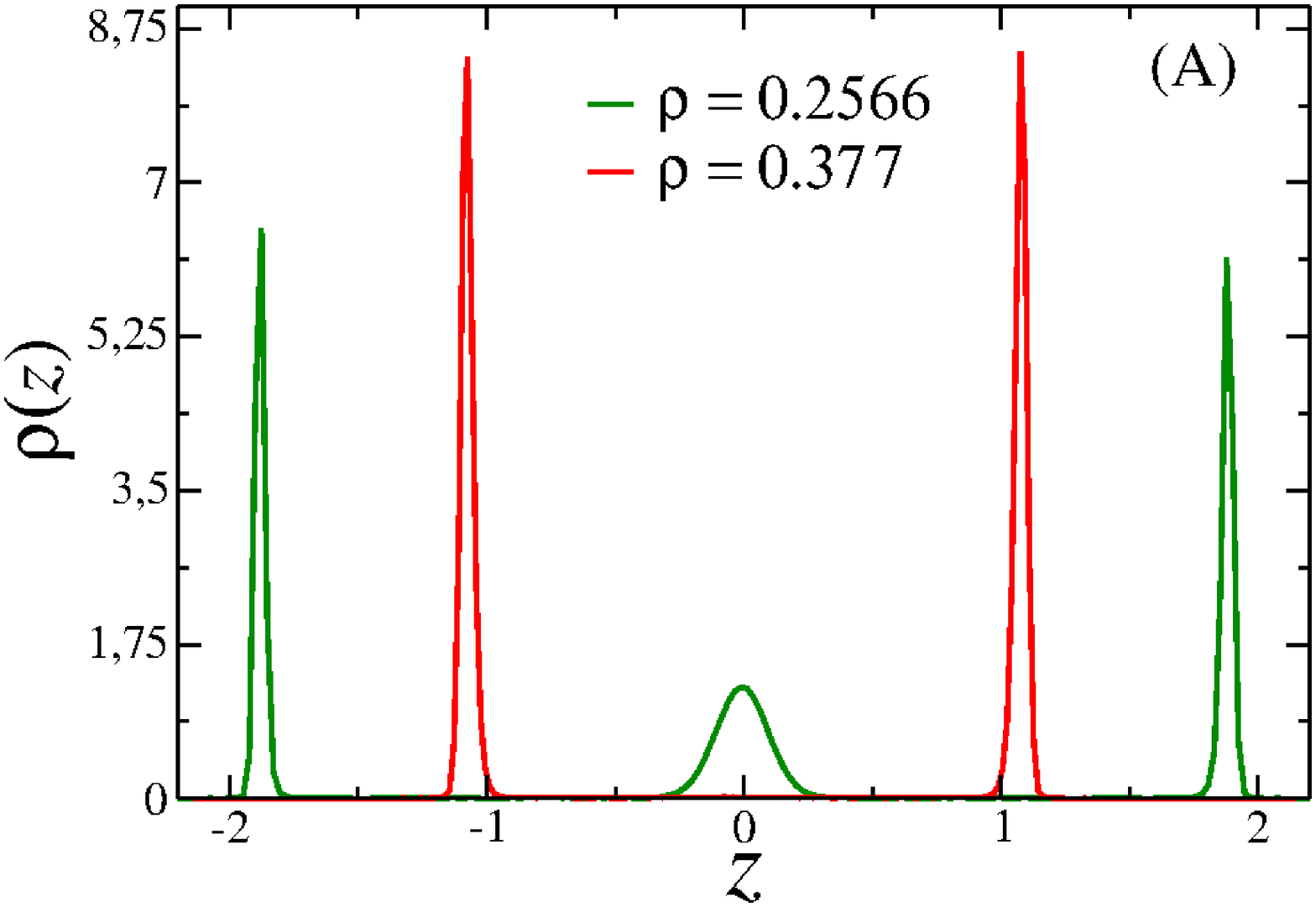}
\includegraphics[width=8cm]{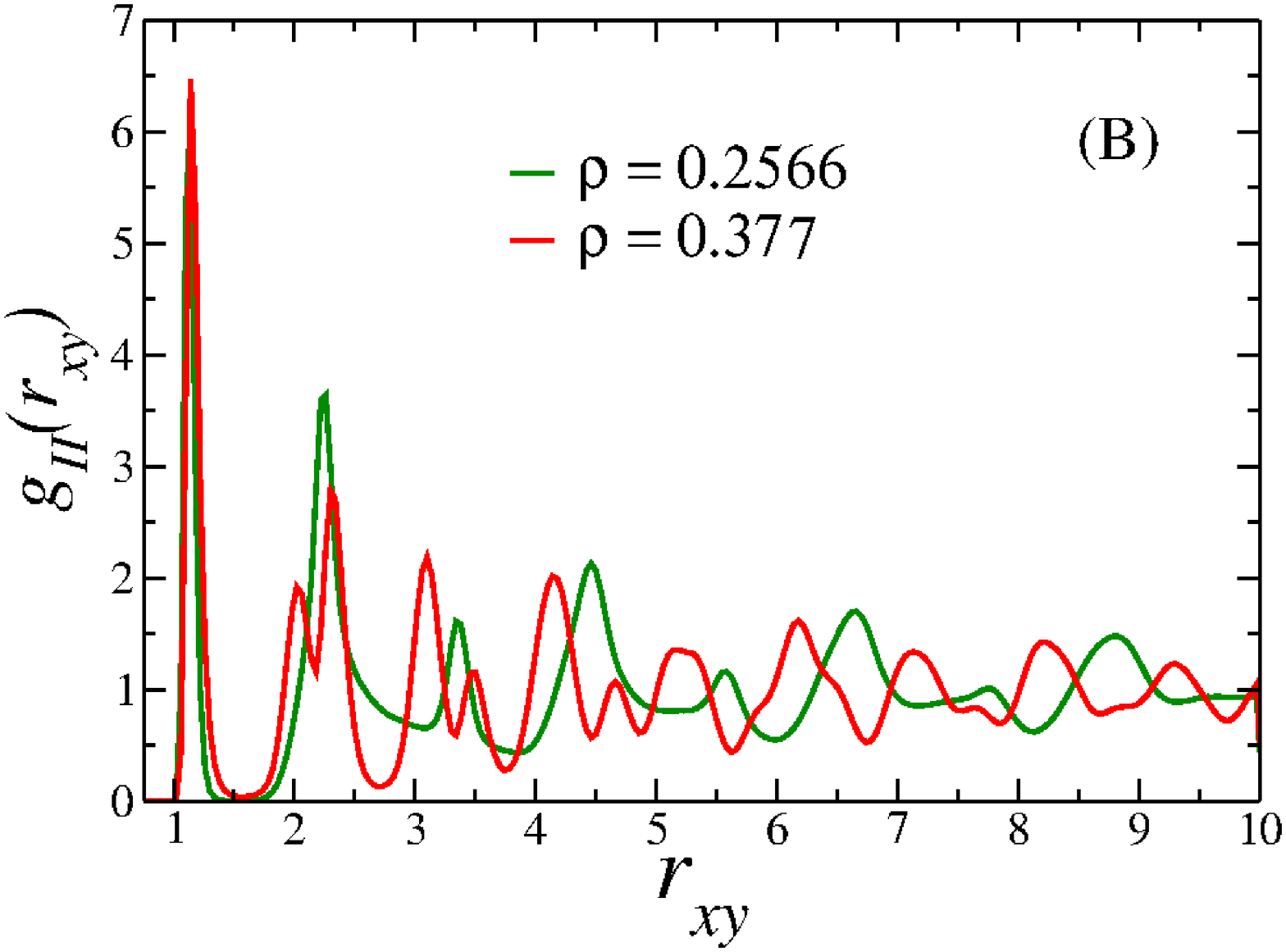}
\includegraphics[width=8cm]{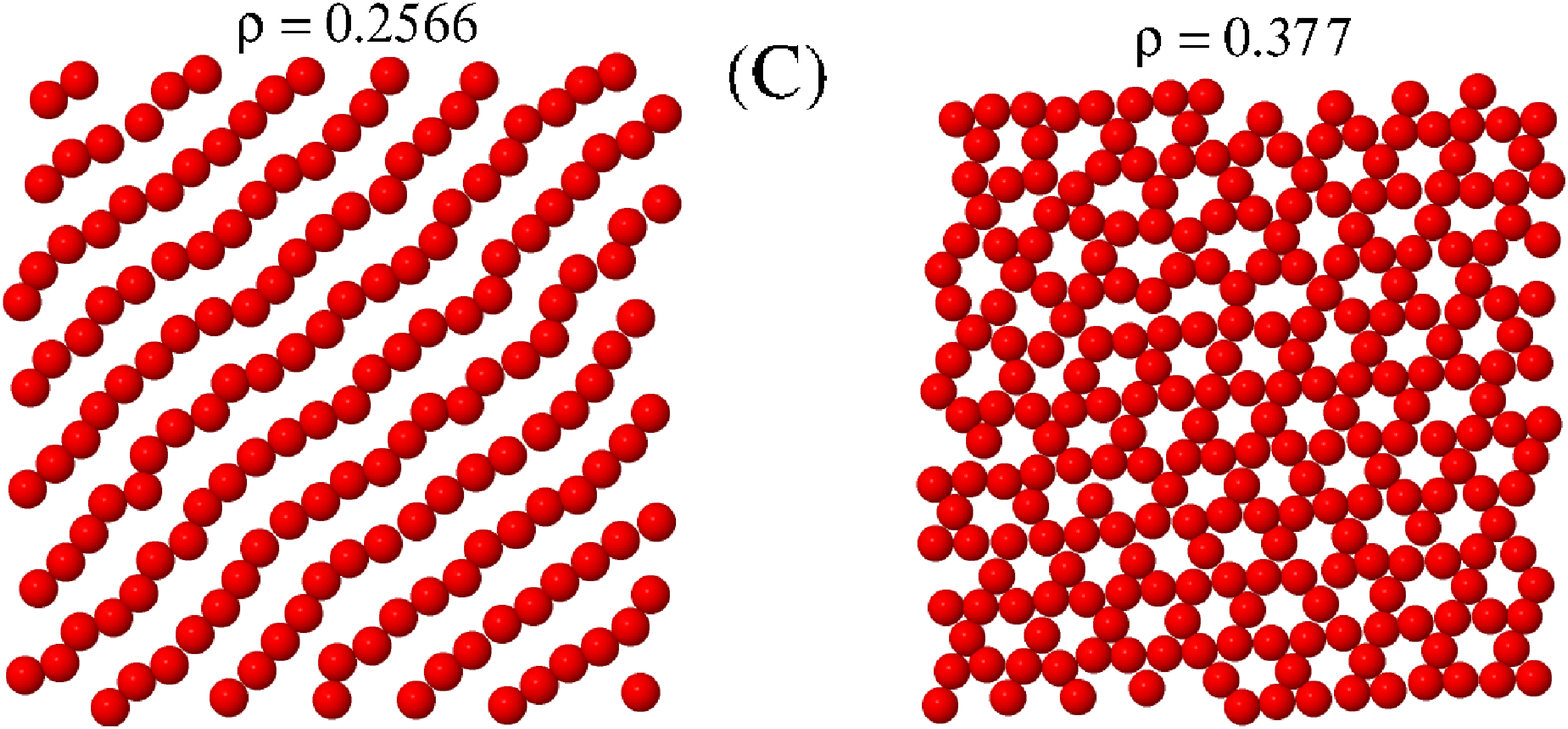}
\end{center}
\caption{Density histogram (A), radial distribution function (B) and snapshots (C) for 
the anomalous fluid confined inside a rigid nanopore at $T = 0.075$
and in the points V and VI from figure~\ref{fig5}.}
\label{fig8}
\end{figure}

Let us first analyze the fluid confined within rigid walls.
The fluid within the walls form layers and as the density
decreases, the number of layers increases.
Figure~\ref{fig6}(A) illustrates that for $T=0.075$
and pressure $p_{z}=0.762$
the system can have five or four layers,
and the density would be  $\rho_I=0.1493$ or $\rho_{II}=0.1687$ respectively. 
Figure~\ref{fig7}(A) shows that  the system can have four  or three layers 
at the same temperature and perpendicular pressure
if the density would be $\rho_{III}=0.18562$ or $\rho_{IV}=0.228$ respectively. 
Finally, the figure~\ref{fig8}(A) has the the system with three or two
layers depending if the density would be $\rho_{V}=0.2566$ or $\rho_{VI}=0.377$, respectively,
with the same value of $p_z$.
These three coexisting regions  at $T=0.075$ are illustrated 
in the temperature versus density
phase diagram of figure~\ref{fig5} as I, II, III, IV, V and VI.

The phase transition observed in the figure~\ref{fig5} 
can be associated 
with the change in the number of layers. 
In order to explore the idea that this transition is 
also associated with changes in the structure of the contact layer,
the radial distribution function $g_{\parallel}(r_{xy})$
of the contact layer was computed. Figure~\ref{fig6}(B)
indicates that  for $T=0.075$, $p_z=0.762$ and
the densities $\rho_I$ or $\rho_{II}$ the 
contact layers exhibit two distinct  structures.
The high peak in the second length scale for the $g_{\parallel}(r_{xy})$
of $\rho_I=0.1493$ and the fact that between the two first peaks the 
radial distribution function is not equal to zero indicates that this
layers is in a liquid-crystal like state. 
The $g_{\parallel}(r_{xy})$ for the density $\rho_{II}=0.1687$ 
shows a very structured liquid as well. It
has a higher first peak when compared with the
peak in the case $\rho_I$. The  $\rho_{II}$ also
has a  displacement in the subsequent peaks what
suggests an additional length scale in the arrangements of the particles.
These two distinct particle arrangement 
are illustrated in the 
snapshots of the contact layer
shown in figure~\ref{fig6}(C).
These pictures confirm the two structures predicted by
the radial distribution function. The
 dimeric arrangement corresponds to the increase 
in the first peak of the $g_{\parallel}(r_{xy})$
for $\rho_{II}$ while the displacement of 
of the other peaks represent the second length.
The systems exhibits a    first-order phase transition
from 
liquid-crystal-like to  
a dimeric structured liquid in the contact layer.

The  $g_{\parallel}(r_{xy})$
for the densities
 $\rho_{III} = 0.18562$ and $\rho_{IV} = 0.228$,  associated with the four to three layers
in figure~\ref{fig7}(A)
respectively,  are
shown  in figure~\ref{fig7}(B). As the density changes from $\rho_{II}$ to $\rho_{III}$ 
the dimeric system becomes continuously  more structured. As
the density increases further, at $\rho_{IV}$ the
system changes discontinuously to an ordered solid structure.
As the snapshots in the figure~\ref{fig7}(C) indicates, the dimers observed in
figure~\ref{fig6}(C) are now forming disordered lines in the density $\rho_{III}$. 
For the higher density $\rho_{IV}$ the line are completely arranged in a ordered structure.
This indicates that the presence of a second first order  transition from a 
structured liquid phase to a solid phase.

The radial distribution functions for densities $\rho_{V} = 0.2566$ and $\rho_{VI} = 0.377$
illustrated  in figure~\ref{fig8}(B) 
show a coexistence of two different highly ordered 
solid-like structures in the contact layer. Analysis using the snapshot
 of the figure~\ref{fig8}(C)
shows that for $\rho_{V}$ the particles form line
that can be in different orientations. More important than this, the snapshots
shows a structural transition from the lined conformation to a honeycomb structure. This
surprising result shows that the third van der Waals loop corresponds to a solid-solid
phase transition in the contact layer.

The three phase transitions at the surface are represented by
the density jumps in the temperature versus density 
phase transition in  figure~\ref{fig5} and by
the van der Waals loops in figure~\ref{fig3} showing that
the instabilities signalized in these graphs are 
related to phase transitions at the fluid wall interface.
The transition between two solid or solid-like phases
usually imply a change in the order parameter  symmetry and, therefore,
can not be modeled by a van der Waals theory. However, 
Daanoun, Tejero and Baus showed that
the van der Waals theory can be extended to solid-solid transitions~\cite{Daanoun94}
in some special cases.  In this context, a number of solid-solid first-order
phase transitions ending in critical points
where found~\cite{Bolhuis94,Bolhuis97,Dijkstra02,Lo97,De97,Ma97},
particularly in 2D systems in which the 
particles interact through a two length
scales potential~\cite{Ja98,Du13,Yo77}. 
Therefore, since the contact layer is a kind of 2D system,
our model falls in the category and is 
not surprising that the surface would exhibit
a solid-solid phase transition.

\begin{figure}[H]
\begin{center}
\includegraphics[width=10cm]{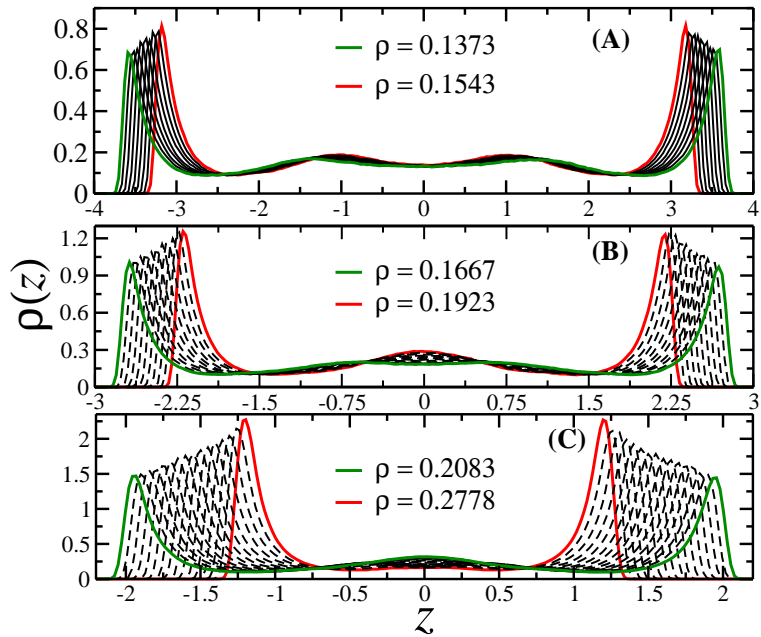}
\end{center}
\caption{Density histogram for the anomalous fluid
confined inside a rigid nanopore at temperatures above the critical point, $T = 1.0$.}
\label{fig9}
\end{figure}
\begin{figure}[H]
\begin{center}
\includegraphics[width=10cm]{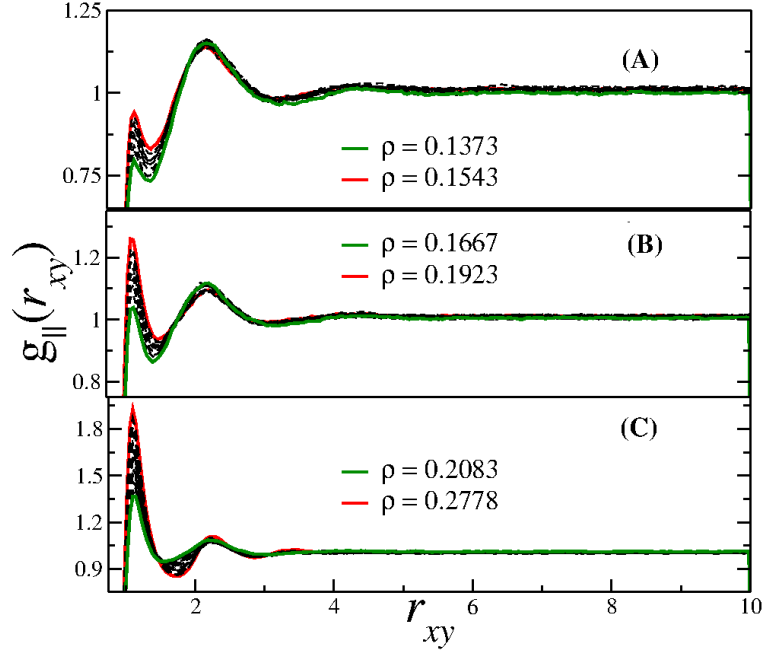}
\end{center}
\caption{$g_{\parallel}(r_{xy})$ for the anomalous fluid
confined inside a rigid nanopore at temperatures above the critical point, $T = 1.0$.}
\label{fig10}
\end{figure}

For temperatures above the critical region the number 
of layers of the anomalous fluid
confined inside a rigid nanopore does not change significantly.
Figures~\ref{fig9}(A), (B) and (C) illustrates
the density across the nanopore for various plates separation,
showing two contact layers and an uniform distribution
inside the pore. The radial distribution function of 
the contact layer, presented in  
figure~\ref{fig10}(A), (B) and (C), 
shows a fluid like behavior for all densities. 
In this way, at high temperatures the layers transition does 
not occur and the structure of the contact layer
do not change, and none phase transition is observed.

\begin{figure}[H]
\begin{center}
\includegraphics[width=10cm]{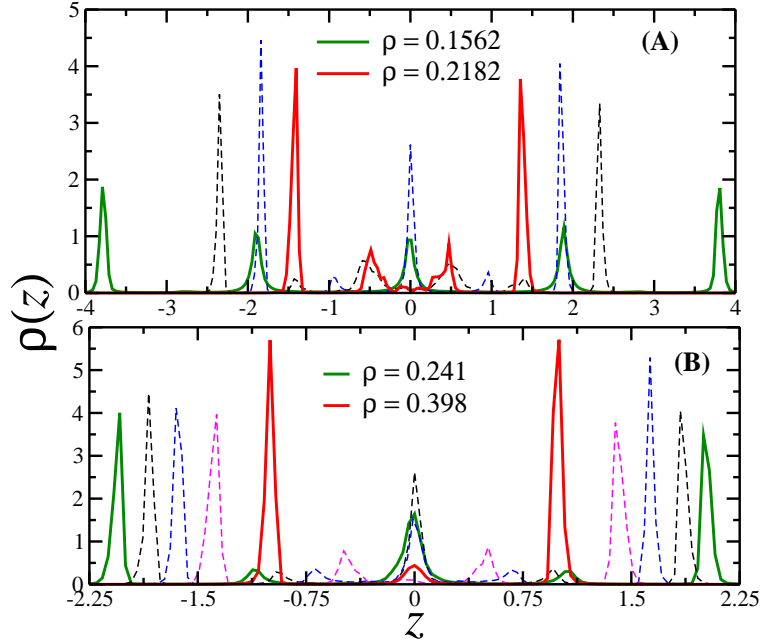}
\end{center}
\caption{Density histogram (left - figures A and B) and $g_{\parallel}(r_{xy})$ (right - figures C and D) for the anomalous fluid
confined inside a flexible nanopore at a low temperature, $T = 0.10$}
\label{fig11}
\end{figure}

\begin{figure}[H]
\begin{center}
\includegraphics[width=10cm]{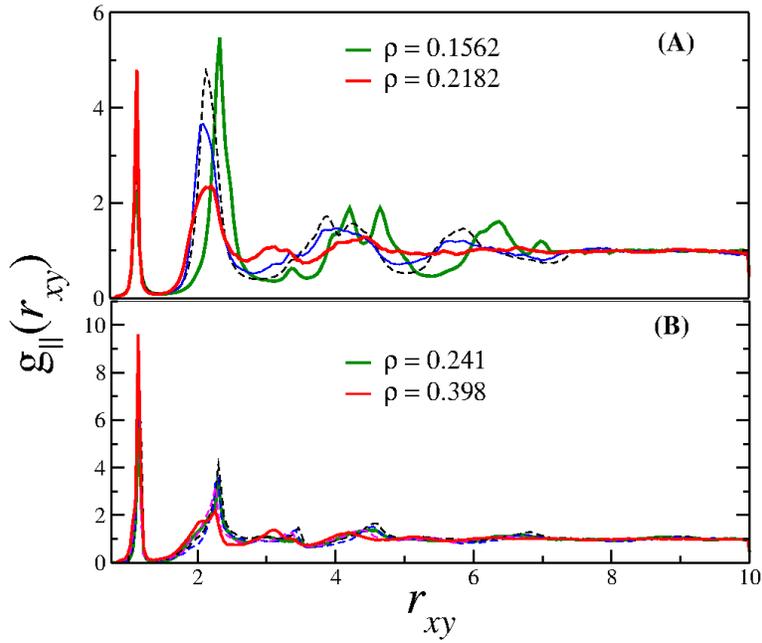}
\end{center}
\caption{$g_{\parallel}(r_{xy})$ (right - figures C and D) for the anomalous fluid
confined inside a flexible nanopore at a low temperature, $T = 0.10$.}
\label{fig12}
\end{figure}
\begin{figure}[H]
\begin{center}
\includegraphics[width=12cm]{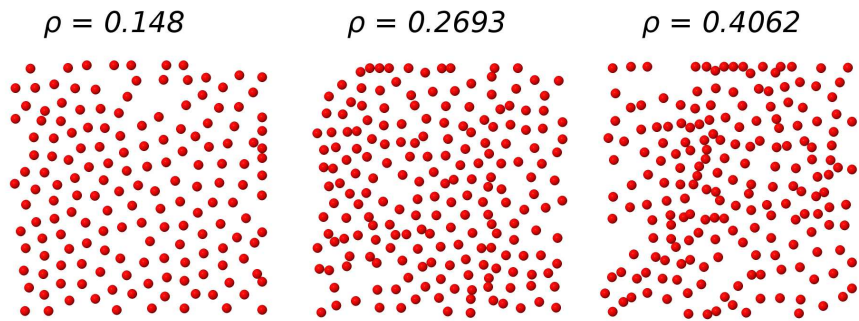}
\end{center}
\caption{Contact layer snapshots in the flexible nanopore case. In all cases, a fluid-like structure was observed.}
\label{fig13}
\end{figure}

Next, we analyze the behavior of the anomalous fluid inside a flexible nanopore.
Figure~\ref{fig11}(A) and (B) shows the 
number of layers for different densities at $T=0.10$. As the density
is increased the number of layers decrease from five to two layers.   
The nanopore flexibility leads also to a distinct behavior in the contact layer 
structure, which is 
strongly affected by the walls movement. Figure~\ref{fig12}(A) and (B)
illustrates the radial distribution function of the contact layer
for various densities. In all the cases the $g_{\parallel}(r_{xy})$
shows a distinct signature of amorphous phase. 
This observation is supported by the  snapshot shown in figure~\ref{fig13}.
The system exhibits a disordered structure similar to the 
amorphous phase. No phase transition is present as already indicated
by the figure~\ref{fig4}(A).

\begin{figure}[H]
\begin{center}
\includegraphics[width=10cm]{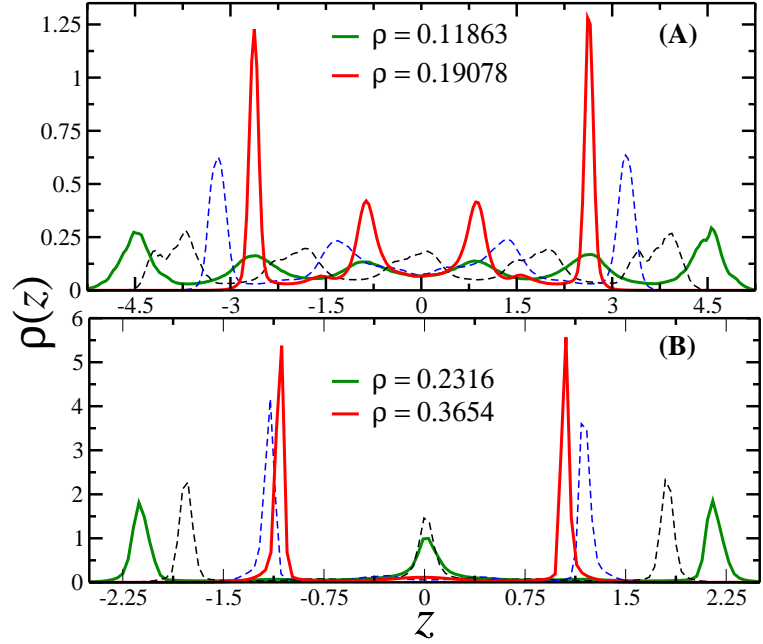}
\end{center}
\caption{Density histogram for the anomalous fluid
confined inside a flexible nanopore at a elevated temperature, $T = 1.0$.}
\label{fig14}
\end{figure}
\begin{figure}[H]
\begin{center}
\includegraphics[width=10cm]{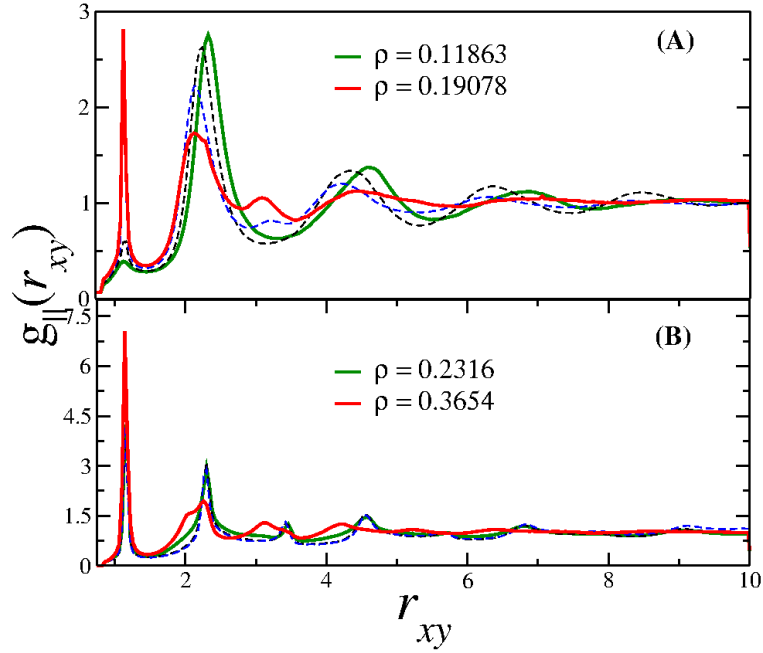}
\end{center}
\caption{$g_{\parallel}(r_{xy})$ for the anomalous fluid
confined inside a flexible nanopore at a elevated temperature, $T = 1.0$.}
\label{fig15}
\end{figure}

Is important to point that the wall flexibility, despite maintaining the contact 
layer in a disordered structure, 
make more difficult to destroy the central layers  at
higher temperatures. At right temperatures, as $T= 1.0$, 
and small density the system shows a bulk-like density profile, as 
shown in figure~\ref{fig14}(A).
But, for slightly higher densities, layering of the fluid is restored
as illustrated in figure~\ref{fig14}(B). This
behavior is distinct from the rigid nanopore case in which 
for any density at high temperatures the layering is
lost. Due to the wall
oscillations the fluid particles can 
assume in the $z$-direction a position that minimizes the energy. And this 
small oscillation, compared with
fixed walls, leads to the layering even for high temperatures, as shown in 
figure~\ref{fig14}(A) and (B). This order in the middle layers 
at high temperatures does not affect the contact layer. Figure~\ref{fig15}(A) and (B)
shows that the  the radial distribution function
of the contact layer is similar to the
amorphous phase.

\section{Conclusion}

We have studied the thermodynamical behavior and the surface phase transition of a 
anomalous fluid confined inside rigid and flexible
nanopores. Our results show that the fluid behavior is strongly affected 
by the confinement properties.
In the rigid nanopore scenario, the $p_{z}\times\rho$ phase diagram
shows the presence of 
three first order phase transitions related with 
structural phase
transitions at the contact layer. 
Due the walls fluctuations in the flexible nanopore case,
no surface phase transition is observed
in the case of non rigid walls. Our results indicates
that the thermodynamic behavior of anomalous fluids such 
as water obtained for rigid carbon nanotubes and solid state nanopores can not
be extrapolated to more flexible walls such as the surface present
in biological systems.

 \section{Acknowledgments}

JRB would like to thanks Prof. A. Diehl from UFPel for the discussions.
We thanks the Brazilian agencies CNPq, INCT-FCx, and Capes for the finantial support.
We also thanks to CEFIC - Centro de F\'{i}sica Computacional of Physics Institute at UFRGS
and the TSSC - Grupo de Teoria e Simula\c{c}\~{a}o em Sistemas Complexos 
at UFPel for the computer clusters.


%

\end{document}